\documentclass[aps,prb,twocolumn,groupedaddress]{revtex4}

\usepackage{graphics} 
\usepackage{graphicx}
\usepackage{dcolumn}
\usepackage{bm}
\everymath{\displaystyle}
\usepackage{amsmath}

\setcitestyle{super}

\begin{document}

\title{Temperature Damping of Magneto-Intersubband  Resistance Oscillations in Magnetically Entangled Subbands} 

\author{Sara Abedi}
\author{Sergey Vitkalov}
\email[Corresponding author: ]{svitkalov@ccny.cuny.edu}
\affiliation{Physics Department, City College of the City University of New York, New York 10031, USA}
\author{A. A. Bykov}
\author{ A. K. Bakarov}
\affiliation{Rzhanov Institute of Semiconductor Physics, Siberian Branch, Russian Academy of Sciences, Novosibirsk, 630090, Russia}

\date{\today}

\begin{abstract} 

Magneto-intersubband resistance oscillations (MISO) of highly mobile 2D electrons  in  symmetric GaAs  quantum wells with two populated subbands  are studied in magnetic fields  tilted from the normal to the 2D electron layer at different temperatures $T$. Decrease of MISO amplitude with temperature increase is observed.  At moderate tilts the temperature  decrease of  MISO amplitude  is consistent with decrease  of Dingle factor due to reduction of  quantum electron lifetime at high temperatures.  At large tilts new regime of strong MISO suppression with the temperature is observed. Proposed model relates this suppression to magnetic entanglement between subbands,  leading to beating in oscillating density of states.  The model yields corresponding temperature damping factor: $A_{MISO}(T)=X/\sinh(X)$, where $X=2\pi^2kT\delta f$ and $\delta f$ is  difference frequency  of  oscillations of density of states in two subbands.  This factor  is in agreement with experiment. Fermi liquid enhancement of MISO amplitude is observed.  

\end{abstract}
 

\maketitle

\section{Introduction}
The orbital quantization of electron trajectories and spectrum  in magnetic fields significantly affects  the electron  transport  in condensed materials\cite{shoenberg1984,ziman,ando1982}.   Shubnikov-de Haas (SdH) resistance oscillations\cite{shoenberg1984} and   Quantum Hall Effect (QHE)\cite{qhe} are  remarkable effects of the  orbital quantization. These effects occur at a temperature, $T$, which is less than the cyclotron energy, $\Delta_c=\hbar \omega_c$, separating  Landau levels. Here $\omega_c$ is the cyclotron frequency. At  high temperatures, $kT> \hbar \omega_c$,  both SdH oscillations and QHE are absent due to  a spectral averaging of the oscillating density of states (DOS)  in the energy interval, $\delta \epsilon \approx kT$, in a vicinity  of Fermi energy, $\epsilon_F$. 

At  the high temperatures, $kT> \hbar \omega_c$, electron systems with multiple populated subbands continue to exhibit quantum resistance oscillations.\cite{coleridge1990,leadley1992,bykov2008a,mamani2008,goran2009,bykov2010} These magneto-inter-subband oscillations (MISO) of the resistance  are due to an alignment between Landau levels from different subbands $i$ and $j$ with corresponding energies $E_i$ and $E_j$ at the bottom of the subbands. Resistance maxima occur at magnetic fields in which the gap between the bottoms of the subbands, $\Delta_{ij}=E_i-E_j$, is a multiple of the Landau level spacing: $\Delta_{ij}=k\cdot\hbar\omega_c$, where $k$ is an integer \cite{magaril1971,polyanovskii1988,raikh1994,averkiev2001,raichev2008}.  At this condition Landau levels of two subbands overlap and  the electron  elastic scattering on  impurities is enhanced due to  the possibility of electron transitions between the overlapped  quantum levels of $i$-th and $j$-th subbands. At magnetic fields corresponding to the condition $\Delta_{ij}=(k+1/2)\cdot\hbar\omega_c$ the intersubband electron scattering is suppressed since the quantum levels of two subbands are misaligned. The  spectral overlap between two subbands oscillates with the magnetic field and   leads to  MISO, which are periodic in the inverse magnetic field. 

Recently we have studied transport properties of  high quality GaAs quantum wells with two populated subbands in a tilted magnetic fields.\cite{mayer2017}    The goals of that study were to detect effects of the spin (Zeeman) splitting on MISO, which has not been seen before as well as to investigate the effect of the spin splitting on quantum positive magnetoresistance (QPMR)\cite{vavilov2004,mamani2009,dietrich2012,mayer2016a}  in a 2D system with two populated subbands . These experiments  have  demonstrated a significant reduction of the QPMR with the application of the in-plane magnetic field, which was in  good agreement with the  modification of the electron spectrum via Zeeman effect with g-factor $g \approx$0.43$\pm$0.07.  MISO also have  a strong reduction of the magnitude with the in-plane magnetic field. However in contrast to the QPMR, the MISO reduction is  found to be predominantly related to a modification of the electron spectrum via a magnetic entanglement  of two subbands, induced by the in-plane magnetic field.\cite{mayer2017} 

In zero magnetic field the electron motion in a quantum well can be separated on two independent parts: the lateral motion along the 2D layer and  the vertical motion (perpendicular to 2D layer), which  is quantized. In a perpendicular magnetic field the lateral motion is also quantized, forming Landau levels, but the lateral and vertical motions are still separable. The eigenstates of the systems can be, therefore,  represented as a product of two wave functions, corresponding to two eigenstates for vertical and lateral motions.  The in-plane magnetic field couples vertical and lateral electron motions making these electron motions to be non-separable or entangled. As a result, in a tilted magnetic field  the eigenstates of the system cannot be presented  as a product of two wave functions, corresponding to lateral and vertical motions  but  are presented as a linear superposition of such products. In this paper we call this effect  magnetic entanglement of two subbands since mathematically the effect is similar to the quantum entanglement of particles in many body physics. 
 
It is important to mention that the Hamiltonian (\ref{ham}), describing the entangled subbands, appears in QED models, where a photon mode/harmonic oscillator, represented in our case by Landau levels, couples to a qubit, represented by two subbands. Such systems have been used  in atomic physics\cite{haroche2013} and quantum optics as well as with superconducting circuits.\cite{fink2010, pla2016} Recently this model was exploited for 2D electrons on the surface of liquid He-4.\cite{yunus2019}

In this paper the temperature dependence of MISO amplitude is studied in a broad range of angles $\theta$ between the magnetic field, ${\bf B}$, and the normal to the 2D layer. At small angles the MISO temperature dependence is controlled by temperature variations of the electron quantum lifetime entering the Dingle factor. At large angles $\theta$ a new regime of the temperature damping of MISO is observed demonstrating an exponentially  strong decrease of MISO magnitude with the temperature. The proposed model relates the observed MISO suppression with the magnetic entanglement of subbands leading to MISO damping factor:  $A_{MISO}(T)=X/\sinh(X)$, where $X=2\pi^2kT\delta f$ and $\delta f$ is  difference frequency  of  oscillations of density of states in two subbands. A comparison with the model reveals   enhancement of MISO magnitude, which has  Fermi liquid origin. 

The paper has the following organization. Section II presents details of experimental setup. Experimental results are presented in section III. In section IV the model leading to MISO is discussed in details. Section V presents  comparison and discussion of experimental results and the model outcomes. Appendix 1 presents cyclotron mass calculations and computations of the parameter X  for magnetically entangled subbands. Appendix B contain details of the derivation of Eq.(\ref{cond_fin}).

\section{Experimental Setup}

Studied GaAs quantum wells were grown by molecular beam epitaxy on a semi-insulating (001) GaAs substrate. The material was fabricated from a selectively doped GaAs single quantum well of width $d$=26 nm sandwiched between AlAs/GaAs superlattice screening barriers.\cite{friedland1996,dmitriev2012,kanter2018,sammon2018,akino2021}  The studied samples were etched in the shape of a Hall bar. The width and the length of the measured part of the samples are $W=50\mu$m and $L=250\mu$m. AuGe eutectic was used to provide electric contacts to the 2D electron gas. Samples were studied at different temperatures, from 5.5 Kelvin to 12.5 Kelvin in magnetic fields up to 7 Tesla applied  at different angle $\theta$ relative to the normal to 2D layers and perpendicular to the applied current.  The angle $\theta$ is evaluated using Hall voltage $V_H = B_\perp/(en_T)$, which is proportional to the perpendicular component, $B_\perp=B\cdot cos(\theta)$, of the total magnetic field ${\bf B}$.     

The total electron density of sample S1, $n_T= (8.0 \pm 0.03)\times 10^{11} cm^{-2}$, was evaluated from the Hall measurements taken  in classically strong magnetic fields \cite{ziman}. An average electron mobility $\mu \approx 72 m^2/Vs$  was obtained from $n_T$ and the zero-field resistivity. An analysis of the periodicity of MISO in the inverse magnetic field yields the gap $\Delta_{12}$=15.15 meV between bottoms of the conducting subbands,  Fermi energy $E_F$=21.83 meV and  electron densities  $n_1$=6.12$\times 10^{11} cm^{-2}$ and  $n_2$=1.87$\times 10^{11} cm^{-2}$ in the  two populated subbands.  Sample S2  has density $n_T\approx 8.0\times 10^{11} cm^{-2}$,  mobility $\mu \approx 100 \times m^2/Vs$ and the gap $\Delta_{12}$=15.10 meV.  Both samples have demonstrated  similar behavior in  magnetic fields. Below we present data for sample S1.

Sample resistance was measured using the four-point probe method. We applied a 133 Hz $ac$ excitation $I_{ac}$=1$\mu$A  through the current contacts and measured the longitudinal (in the direction of the electric current, $x$-direction) and Hall $ac$ (along $y$-direction) voltages ($V^{ac}_{xx}$ and $V^{ac}_H$) using two lock-in amplifiers with 10M$\Omega$ input impedance.  The measurements were done in the linear regime in which the voltages are proportional to the applied current. 

\section{Experimental Results}

\begin{figure}[t]
\vskip -0 cm
\includegraphics[width=\columnwidth]{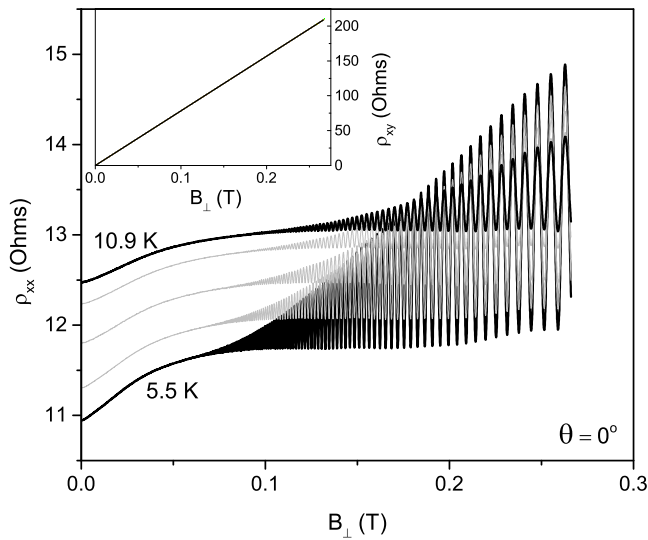}
\caption{Dependencies of the dissipative resistivity of 2D electrons, $\rho_{xx}$, on perpendicular magnetic field taken at different temperatures: from bottom to top  T=5.5, 6.9, 8.5, 10.1, and 10.9K.  The inserts shows the Hall resistivity, $\rho_{xy}$, in a perpendicular magnetic field at the same set of temperatures as in the main plot.  Angle $\theta=0^0$. }
\label{miso_exp1}
\end{figure}

Figure \ref{miso_exp1} shows dependencies of the dissipative resistivity of 2D electrons on the perpendicular magnetic field $B_{\perp}$, taken at different temperatures T and the angle $\theta = 0^0$  between the direction of the  magnetic field ${\bf B}$ and the normal to the 2D layer. At  $\theta=0^0$ two subbands are disentangled. At T = 5.5 K and small magnetic field ($B_{\perp} <$  0.05 T), the curve demonstrates an increase related to classical magnetoresistivity.\cite{ziman,mayer2017} At higher magnetic fields, $B_{\perp} >$ 0.08 T, the resistivity starts to oscillate with progressively larger magnitude at higher field. These oscillations are  MISO. MISO maxima correspond to the condition
\begin{equation}
\Delta_{12} = k \hbar \omega_c
\label{miso_eq}
\end{equation}
,  where $\Delta_{12} = E_2- E_1$ is the energy difference between bottoms of two occupied subbands and the index $k$ is a positive integer.\cite{raikh1994,raichev2008}

The temperature significantly affects  the MISO magnitude. At temperature 10.9K the MISO magnitude is substantially smaller the one at T=5.5K. Furthermore at a higher temperature the oscillations starts at a higher  magnetic field. Both effects are a result of an increase of the quantum scattering rate of electrons at higher temperature due to the enhancement of electron-electron scattering.\cite{mamani2008,goran2009,dietrich2012}  This rate enters the Dingle factor, affecting strongly MISO magnitude (see below Eq.(\ref{cond_fin})).   The insert to Fig.\ref{miso_exp1} shows the Hall resistivity at different temperatures. The insert indicates that the Hall resistivity and, thus,  the total electron density in the system  are not affected by the temperature.

\begin{figure}[t]
\vskip -0 cm
\includegraphics[width=\columnwidth]{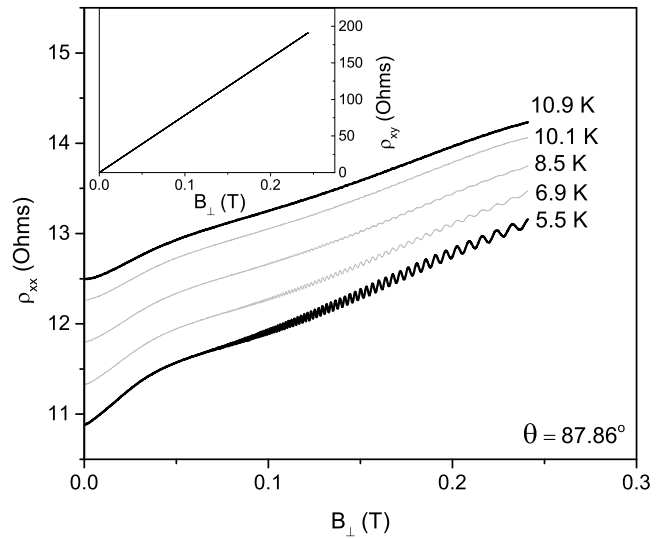}
\caption{Dependencies of the dissipative resistivity of 2D electrons, $\rho_{xx}$, on perpendicular magnetic field taken  at different temperatures: from bottom to top  T=5.5, 6.9, 8.5, 10.1, and 10.9K.  The inserts shows the Hall resistivity, $\rho_{xy}$, in a perpendicular magnetic field at the same set of temperatures as in the main plot. Angle $\theta=87.86^0$. }
\label{miso_exp2}
\end{figure}

Figure \ref{miso_exp2} shows dependencies of the dissipative resistivity of 2D electrons on the perpendicular magnetic field $B_{\perp}$, taken at different temperatures T but at the angle $\theta=87.86^0$. At  $\theta=87.86^0$ two subbands are  entangled by the in-plane magnetic field.  At T = 5.5 K and small magnetic field ($B_{\perp} <$  0.05 T), the curve continue to demonstrate an increase related to classical magnetoresistivity.\cite{ziman,mayer2017} At higher magnetic fields, $B_{\perp} >$ 0.08 T, the resistivity starts to oscillate but with  a magnitude, which is  significantly smaller than the one shown in Fig.\ref{miso_exp1} for disentangled subbands. The insert to the figure indicates that the Hall resistivity and the total electron density, $n_T$, are still  temperature independent and stays the same as for disentangled subbands.

\begin{figure}[b]
\vskip -0 cm
\includegraphics[width=\columnwidth]{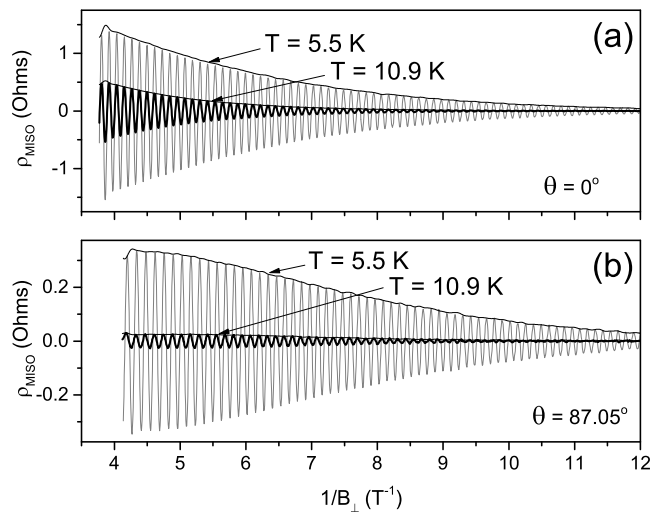}
\caption{Oscillating content  of magnetoresitivity $\rho_{xx}$  at two  different temperatures as labeled. (a) disentangled subbands at angle $\theta=0^0$;  (b) entangled subbands at angle $\theta=87.05^0$.}
\label{miso_exp3}
\end{figure}

To facilitate the analysis of the oscillating content, the monotonic background $\rho^{b}_{xx}$, obtained by an averaging of the oscillations in reciprocal perpendicular magnetic fields, is removed from the magnetoresistivity $\rho_{xx}$($B_{\perp}$). Figure \ref{miso_exp3} presents the remaining oscillating content of the magnetoresistivistity, $\rho_{MISO}$,  as a function of the reciprocal perpendicular magnetic field $B^{-1}_{\perp}$ for two temperatures as labeled. The thin solid lines indicate  envelopes of the oscillating content used in the analysis below.

For disentangled subbands  Figure \ref{miso_exp3}(a) demonstrates  that at the high temperature $T$=10.9 K the MISO magnitude is smaller than the one at $T$=5.5 K.  An analysis of the MISO envelope indicates that  the MISO magnitude decreases  exponentially with $1/B_\perp$ at a  small $1/B_\perp$.  The rate of the exponential  decrease is stronger at  the higher temperature. Both the thermal suppression of MISO and the enhancement of the MISO reduction with $1/B_\perp$   result  from  the increase  of the quantum scattering rate of 2D electrons, $1/\tau_{q}$, due to the increase of electron-electron scattering at high temperatures. 

Figure \ref{miso_exp3}(b)  demonstrates the dependence of  MISO on $1/B_\perp$ for the magnetically entangled subbands  at $\theta =$ 87.05 degrees.  The decrease of MISO magnitude with $1/B_\perp$ is  different from the exponential decrease of the disentangled subbands. The magnetic field dependence tends to saturate at small $1/B_\perp$ in contrast to  the one shown in  Figure \ref{miso_exp3}(a).  For the entangled subbands the MISO magnitude is significantly reduced.  Furthermore a rough analysis indicates  that  the relative decrease of the MISO magnitude with the temperature is substantially stronger than the one for disentangled subbands. In particular, at $1/B_\perp=$ 5 (1/T) for the disentangled subbands the ratio between MISO magnitudes at $T_1$=5.5K   and $T_2$=10.9 K is close to 3, while for the entangled subbands the ratio is larger and close to 10.   

\begin{figure}[t]
\vskip -0 cm
\includegraphics[width=\columnwidth]{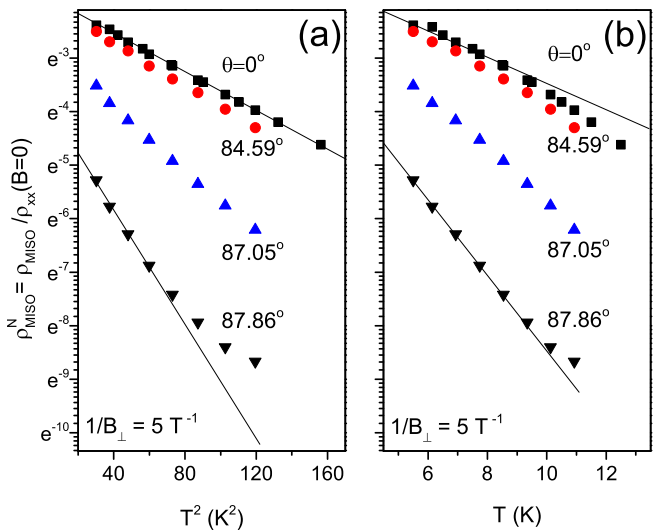}
\caption{Temperature dependence of normalized amplitude of MISO, $\rho^{N}_{MISO}=\rho_{MISO}/\rho_{xx}({\bf B}=0)$ at $B^{-1}_{\perp}$ = 5 (1/T). (a) The dependence is plotted vs  $T^2$; (b) the dependence is plotted vs  $T$.
}
\label{miso_exp4}
\end{figure}

Figure \ref{miso_exp4} presents an evolution of the temperature dependence of the MISO magnitude with the angle $\theta$ at  fixed $B_\perp$=5 (1/T). Figure \ref{miso_exp4}(a) shows the  dependence of the normalized MISO magnitude on $T^2$. At a small subbands entanglement   ($\theta=0^0$ and $84.59^0$)  the MISO magntitude drops exponentially with $T^2$ in a good agreement  with the  solid straight line presenting the $T^2$ exponential decrease at $\theta=0^0$. At  larger angles ($\theta =87.05^0$ and $87.86^0$) the MISO drop becomes  stronger and deviates from the $T^2$ dependence. 

 In Figure \ref{miso_exp4}(b) the symbols present the  dependence of normalized MISO amplitude on  temperature $T$. The solid straight lines demonstrate the exponential decrease with $T$.  At  small subbands entanglement   ($\theta$=0; 84.59 degrees)  the MISO magntitude does not decrease exponentially with $T$.  The dependence deviates considerably from  the  solid straight line.  In contrast  at the largest angle (87.86 degrees) the MISO reduction is consistent with the exponential decrease with $T$ and follows the solid straight line. Thus, Figure  \ref{miso_exp4} shows that the decrease of MISO amplitude with  temperature  is qualitatively different for the entangled subbands, indicating a new mechanism leading to the MISO damping.  This new regime of thermal MISO damping  is analyzed below within  a model, taking into account the  magnetic entanglement of 2D subbands.

\section{Model of Quantum Electron Transport}

 In perpendicular magnetic fields (at $\theta=$0$^o$) a  microscopic theory of MISO  is presented in papers Ref.[\cite{raikh1994,averkiev2001,raichev2008}].  In this theory the electron spectra of two subbands evolve in magnetic fields quite independently. The reason is that at $\theta=$0$^o$ the lateral (in the 2D layer) and vertical (perpendicular to the layer) electron motions are separable and do not affect each other.  In a tilted magnetic field  there is a  component of the field, $B_\parallel$, which is parallel to the  2D conducting layer. This parallel component couples  the lateral  and vertical  electron motions and  electron spectra of two subbands become to be magnetically entangled.  A MISO model, which takes into account this magnetic entanglement between two subbands, is proposed recently. The model demonstrates  significant decrease of MISO amplitude with the magnetic field tilt. \cite{mayer2017}  A comparison with corresponding experiments indicates that the magnetic entanglement between subbands is the dominant mechanism leading to the angular  decrease of the MISO amplitude in GaAs quantum wells.  Zeeman spin splitting  is found to provide a sub-leading contribution to the effect. \cite{mayer2017} 

Below  this model is used to analyze the temperature dependence of the MISO amplitude in tilted magnetic fields. The Zeeman effect is ignored. The analysis reveals  an $universal$  temperature dependent factor, which controls the MISO amplitude in magnetically entangled subbands. The amplitude reduction is found to be exponential with the temperature in the regime of a strong magnetic entanglement.   In many respects the physics of  this additional temperature factor is similar to the one for  SdH oscillations.  The obtained factor describes general MISO property.

\subsection{Spectrum in tilted magnetic field}

Let  2D electrons propagate along  $xy$-plane and the $z$-axes is perpendicular to  the plane. In quantum wells  the spatial subbands are the result of quantization of the electron wave function in the $z$-direction.   Index $i$=1(2) labels the low (high) subband with the energy $E_1$($E_2$) at the bottom of the subband.  The subband separation is $\Delta_{12}=E_2-E_1$.  

With no in-plane magnetic field applied the spatial subbands are coupled to each other via elastic scattering. An in-plane magnetic field, $B_\parallel$,  provides an additional coupling via Lorentz force coming from the last term of the Hamiltonian $H$ presented by Eq.(\ref{ham}). This additional $B_\parallel$-coupling preserves the degeneracy of the quantum levels but induces variations of  the electron spectrum, which, due to the relativistic origin of  the Lorentz force,  are dependent on the energy (velocity). These spectrum variations destroy the complete spectral overlap between Landau levels from different subbands, existing at zero in-plane magnetic field.  This leads to the angular decrease of the MISO amplitude. \cite{mayer2017} Below  we investigate how this  decrease depends on the temperature following to the developed approach.\cite{mayer2017}   

To estimate the effect  the electron spectrum of an ideal two subband system without  impurity scattering is computed numerically in a titled magnetic field. The impurity scattering is introduced then by a broadening  of the bare quantum levels using  Gaussian shape of the DOS with the preserved level degeneracy.  
 
 We consider a quantum well of a width $d$ in $z$-direction formed by a rectangular electrostatic potential  $V(z)$ with  infinitely high walls and placed in a titled magnetic field $ {\bf B}=(-B_\parallel, 0, B_\perp)$. Electrons are described by the Hamiltonian\cite{mayer2017}:
\begin{equation}
H=\frac{\hbar^2 k_x^2}{2m_0}+\frac{e^2B_\perp^2}{2m_0}x^2+\frac{\hbar^2 k_z^2}{2m_0}+V(z) 
+\frac{e^2B_\parallel^2}{2m_0}z^2+\frac{e^2B_\perp B_\parallel}{m_0}xz,\\
\label{ham}
\end{equation}
where $m_0$ is electron band mass.  To obtain Eq.(\ref{ham}) we have used the  gauge (0,$B_\perp x+  B_\parallel z$,0) of the vector potential  and applied the transformation $x \rightarrow x-\hbar k_y/eB_\perp$.  

The first four terms of the Hamiltonian describe the  2D electron system  in a perpendicular magnetic field.  The corresponding eigenfunctions of the system are $\vert N,\xi\rangle$, where $N$=0,1,2.. presents  $N$-$th$ Landau level (the lateral quantization) and $\xi=S, AS$ describes the symmetric (S) and antisymmetric (AS) configurations of the wave function in the $z$-direction (vertical quantization):  $\vert N,S\rangle=\vert N\rangle (2/d)^{1/2}\cos(\pi z/d)$ and  $\vert N,AS\rangle=\vert N\rangle (2/d)^{1/2}\sin(2\pi z/d)$. 

Using  functions $\vert N,\xi\rangle$ as the basis set , one can present the Hamiltonian in  matrix form.  The matrix contains four  matrix blocks: $\hat H=(\hat E^S,\hat T; \hat T, \hat E^{AS})$, where the semicolon separates rows.The diagonal matrices, $\hat E^S$ and $\hat E^{AS}$, represent  energy of the symmetric and antisymmetric wave functions  in different orbital states $N$:    
\begin{equation}
\begin{split}
&E^{S}_{mn}=\delta_{mn}[\hbar \omega_c((n-1)+\frac{1}{2}) +\frac{e^2B_\parallel^2d^2[\frac{1}{12}-\frac{1}{2\pi^2}]}{2m_0}] \\
&E^{AS}_{mn}=\delta_{mn}[\hbar \omega_c((n-1)+\frac{1}{2}) +\Delta_{12} +\frac{e^2B_\parallel^2d^2[\frac{1}{12}-\frac{1}{8\pi^2}]}{2m_0}]
\end{split}
\label{diag}
\end{equation}
where $\Delta_{12}=E_2-E_1$ is the energy difference between bottoms of two spatial subbands and  indexes $m$=1,2...$N_{max}$ and $n$=1,2...$N_{max}$ numerate rows and columns of the matrix correspondingly. These indexes are related to the orbital number $N$: $n,m=N+1$, since the orbital number $N=0,1,2..$. In  numerical computations the maximum number $N_{max}$ is chosen  to be about twice larger than the orbital number $N_F$ corresponding to Fermi energy $\epsilon_F$. Further increase of $N_{max}$ shows a very small (within 1\%) deviation from the dependencies obtained at $N_{max}\approx 2N_F$. This also indicates that the contributions of the third and higher spatial subbands with a higher energy can be ignored in the spectrum computation. It supports the  two subband approximation used in the paper.

The first term in Eq.(\ref{diag}) describes the orbital quantization of electron motion.    The last term in Eq.(\ref{diag}) describes diamagnetic shift of the quantum levels and relates to the fifth term in Eq.(\ref{ham}). In the basis set $\vert N,\xi\rangle$ the diamagnetic  term is proportional to $\langle\xi\vert z^2\vert \xi\rangle$. The diamagnetic terms do not depend on $N$. The diamagnetic terms lead to an increase of the gap, $E_g$,  between bottoms of subbands with the in-plane magnetic field:    
\begin{equation}
E_g(B_\parallel)= \Delta_{12}+\frac{3}{16\pi^2} \frac{e^2B_\parallel^2 d^2}{m_0}  
\label{gap}
\end{equation}

The off-diagonal matrix $\hat T$ is related to the last term in Eq.(\ref{ham}). This matrix mixes symmetric and antisymmetric states. Since $x=l_{B\perp}(a^*+a )/\sqrt2$ works as the raising $a^*$ and lowering $a$ operators of the Landau orbits, the last term in Eq.(\ref{ham}) couples Landau levels  with   orbital numbers different by one. Here  $l_{B\perp}=(\hbar/eB_\perp)^{1/2}$ is the  magnetic length in $B_\perp$. As a result, for $n>m$ the matrix element $T_{mn}$   between states $\vert N, S\rangle$ and $\vert N+1, AS\rangle$ is   
\begin{equation}
\begin{split}
T_{mn}&=\delta_{m+1,n}\frac{e^2B_\parallel B_\perp l_{B\perp}}{m_0}\langle N\vert \frac{a^*+a}{\sqrt2}\vert N+1\rangle \langle S\vert z\vert AS\rangle\\
&=\delta_{m+1,n}\hbar \omega_c\Big[ \frac{16B_\parallel d}{9\pi^2B_\perp l_{B\perp}}\Big](n/2)^{1/2}
\end{split}
\label{offdiag}
\end{equation}
The matrix $\hat T$ is  a symmetric matrix: $T_{mn}=T_{nm}$. 

The Hamiltonian $\hat H$ is diagonalized numerically at different magnetic fields $B_\perp$ and $B_\parallel$. To analyze the spectrum the obtained eigenvalues of the Hamiltonian are numerated in ascending order using positive integer index $l$=1,2...., which is named below as Landau level index.

\begin{figure}[t]
\vskip -0 cm
\includegraphics[width=\columnwidth]{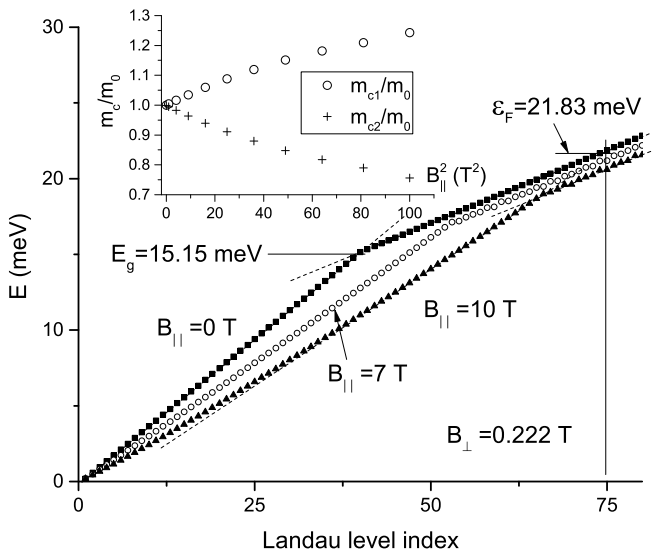}
\caption{Dependence of the energy of Landau levels, counted from the bottom of the lowest subband in GaAs quantum well of width $d$=27 nm,  on Landau level index, $l$,  at different in-plane magnetic fields as labeled. Each symbol corresponds to a Landau level. Kinks in the dependencies occur at the energy corresponding to  the bottom of the second subband, $E_g$. Decrease of the slope of the dependencies at $\epsilon< E_g$ with $B_\parallel$ indicates increase of the cyclotron mass $m_{c1}$ in the first subband. The independence of the slope on $B_\parallel$ at $\epsilon > E_g$ suggests decrease of the mass $m_{c2}$ in the second subband with $B_\parallel$. Vertical line at $l$=75 marks the last populated Landau level in the studied system. $B_\perp$=0.222T. Insert shows divergence of cyclotron masses in two subbands with the in-plane magnetic field.  }

\label{spectrum}
\end{figure}

Figure \ref{spectrum} presents a dependence of the Landau level energy, counted from the bottom of the first subband,  on the index $l$ for different parallel magnetic fields as labeled. In the figure each symbol corresponds to a Landau level.  At $B_\parallel$=0 T and $\epsilon<E_g=\Delta_{12}$ the quantum levels correspond to the first subband. These levels are   evenly separated by the cyclotron energy $\Delta_c=\hbar \omega_c$, forming a straight line. The slope of this line is inversely proportional to the  electron mass, $m_0$,  since $\Delta_c \sim 1/m_0$. The slope is also inversely proportional to the density of states (DOS) since DOS $\sim m_0$ for 2D parabolic bands.   At   $\epsilon > E_g$ the slope of the straight line is abruptly reduced by factor two. This  results from the contribution of the second subband to the total density of states, which starts at   $\epsilon > E_g$. Since the mass in the second subband, $m_0$,  is the same the  total DOS is doubled and the slope is reduced by factor two. The transition between these two straight lines occurs at   $\epsilon = E_g$ and corresponds to the energy of the bottom of the second subband $E_2$.   

At $B_\parallel$=7 T and $\epsilon<\Delta_{12}$ the electron spectrum is different. At the same index $l$ the Landau levels of the first subband have a lower energy indicating an $increase$ of the cyclotron mass in the subband: $m_{c1} > m_0$. This is the effect of the entanglement between subbands, induced by the in-plane magnetic field: the eigenstate $\Psi_l$ of electron performing a cyclotron motion in the tilted magnetic field is now a linear superposition  of the  symmetric $\vert  N, S\rangle$ and antisymmetric states  $\vert  N+n,AS\rangle$ of the Hamiltonian (Eq.(\ref{ham})) at $B_\parallel$=0T.   Although at $B_\parallel$=7 T the open symbols form  apparent straight line,  an analysis indicates  deviations of the data from the linear dependence revealing a non-parabolicity of the spectrum. To simplify the presentation  we neglect  these deviations and approximate the spectrum by a straight line. In other words we consider the spectrum to be parabolic.  Similar to the spectrum at $B_\parallel$=0T the straight line changes its slope due to the  contribution of the second subband to the DOS. The slope change occurs  at a higher energy, $E_g$: $E_g>\Delta_{12}$ due to contribution of the diamagnetic terms to the gap (see Eq.(\ref{gap})). Within  accuracy of one percent  the changed slope coincides with the slope obtained at $B_\parallel$=0T at $\epsilon >\Delta_{12}$. This indicates that at $\epsilon >E_g$ the total density of states is preserved and, therefore, the effective mass in the second subband is reduced by the in-plane field $B_\parallel$: $m_{c2} < m_0$, since $m_{c1}+m_{c2}=2m_0 \sim$ total DOS at high energies. Progressively stronger variations of the masses are seen  at higher in-plane field $B_\parallel$=10T. 

The insert to the figure shows relative variations of the cyclotron masses  in the two subbands induced by the in-plane magnetic field. The insert demonstrates that at small  in-plane magnetic fields the mass divergence is proportional to the square of the field. An analysis  of the two subband model in a small  in-plane magnetic field, given in Appendix A,  provides further support to the presented  interpretation of the electron spectra.  

The insert in Figure \ref{miso_exp2} presents the Hall resistance taken at large tilt: $\theta=86.87$ degrees. The data  indicates that the Hall coefficient, $R_H=1/en_T$, which is the slope of the shown line, does not depend on the in-plane magnetic field. This suggests that the total density $n_T=n_1+n_2$ and,  thus, the electron population of Landau levels at  fixed $B_\perp$: $l_p \approx n_T/n_0$  do not depend on $B_\parallel$. Here $n_0=eB_\perp /(\pi \hbar)$ is the degeneracy of  Landau level (including the spin degeneracy) and $l_p$ is the index of the highest populated level. In Figure \ref{spectrum} the vertical line at $l$=75 marks the highest populated Landau level at $B_\perp$=0.222 T in the studied sample. At a fixed electron density (electron population) the increase of the electron mass $m_{c1}$ drives the Fermi energy, $E_F$, down, while the increase of the energy gap $E_g$ between the subbands moves the Fermi energy up.  An interplay between these  two effects results in a weak decrease of the Fermi energy with the in-plane magnetic field in the studied system.       

\begin{figure}[t]
\vskip -0 cm
\includegraphics[width=\columnwidth]{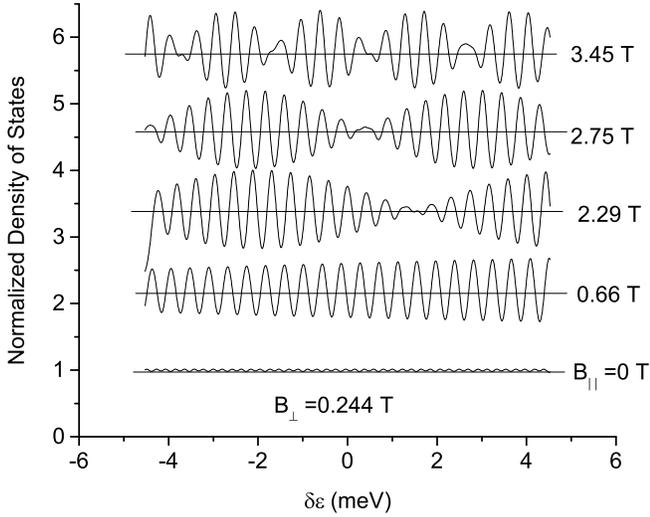}
\caption{Energy dependence of the normalized density of states in the vicinity of Fermi energy: $\delta \epsilon =\epsilon -\epsilon_F$ in quantum well of width $d$=33 nm with two populated subbands, placed in perpendicular magnetic field $B_\perp$=0.244 T and in-plane magnetic fields $B_\parallel$ as labeled.  At $B_\parallel >$0 the dependencies, shifted up for clarity, demonstrate beating pattern. The beating is related to the cyclotron mass divergence presented in the insert to Figure \ref{spectrum}. Quantum scattering time $\tau_q^{(1)}=\tau_q^{(2)}=4 ps$. }
\label{beating}
\end{figure}

The presented  analysis above indicates that in tilted magnetic fields the cyclotron masses in two subbands are different:  $m_{c1} > m_{c2}$. Different cyclotron masses lead to different frequencies  of the DOS oscillations, induced by  the orbital quantization in the energy space. Namely, in the first subband  the DOS  $\nu_1(\epsilon)$ oscillates at frequency $f_1=1/\hbar \omega_{c1} \sim m_{c1}$, while in the second subband the DOS, $\nu_2(\epsilon)$ oscillates at frequency $f_2=1/\hbar \omega_{c2} \sim m_{c2}$, where $\omega_{ci}$ is the cyclotron frequency in $i$-th subband. Thus at the same $B_\perp$ the frequency $f_1$ is higher than $f_2$ since $m_{c1} > m_{c2}$. The difference between frequencies results in a beating of the total DOS oscillations in the energy space as shown in Figure \ref{beating}.
 
 Figure  \ref{beating} demonstrates the total DOS in  a vicinity of Fermi energy: $\delta \epsilon=\epsilon-\epsilon_F$ at fixed perpendicular magnetic field $B_\perp$=0.244 T and different in-plane magnetic field $B_\parallel$ as labeled. The DOS is evaluated via numerical diagonalization  of Hamiltonian (\ref{ham}) and consecutive  broadening of the Landau levels. To demonstrate the DOS beating clearly we use the same quantum scattering time for both subbands $\tau_q^{(1)}=\tau_q^{(2)}$=4 ps.  The obtained  DOS oscillations are well described by an interference of two cosine functions.  At $B_\parallel$=0T  the DOS oscillations are significantly suppressed. This suppression is due to a destructive interference of the DOS oscillations  in two subbands oscillating in anti-phase. This $\pi$-phase shift between the DOS oscillations leads to a MISO minimum, while two in-phase DOS oscillations  should interfere constructively  and lead to a MISO maximum (not shown). A noticeable property of the pattern is that the destructive interference at $B_\parallel$=0T does not depend on the energy. This property is tightly related to the fact that  the DOS  oscillates at the same frequency $f=1/\hbar \omega_c$ in both subbands at $B_\parallel$=0T. 

The DOS oscillations at $B_\parallel$=0.66 T present an example of a partially constructive interference. A noticeable property of these oscillations is an increase of the amplitude of the oscillations with the energy. This property is due to the fact that, in contrast to the DOS interference at  $B_\parallel$=0T, the frequencies of two DOS oscillations at $B_\parallel$=0.66 T are different: $f_1>f_2$. Thus, the interference  pattern between these oscillations depends on the energy,  exhibiting  the beating. The DOS oscillations at $B_\parallel$=2.29, 2.75 and 3.45 T demonstrate the beating pattern with progressively shorter beating periods. The decrease of the beating period or increase of the beating frequency, $f_b$,    is related to the increase of the  difference frequency $\delta f=f_1-f_2=2f_b$  with $B_\parallel$. This increase is due to the mass divergence, shown in the insert to Figure \ref{spectrum}, since $f_i \sim m_i$.  

Below we explain qualitatively why the DOS beating leads to a temperature damping of MISO. More detailed consideration is given in the next section. The electron conductivity is determined by electrons in the $kT$ vicinity of the Fermi energy $\epsilon_F$.\cite{ziman}  The MISO amplitude is determined by the  square of the amplitude of the DOS oscillations averaged within the $kT$ interval \cite{raikh1994, raichev2008}.  Let's assume that the energy interval $kT$ is much less than the beating period ($\sim 1/\delta f$): $kT \delta f \ll 1$.  At this condition the MISO minimum (maximum) occurs when a node (anti-node) of the beating pattern is located in the $kT$ vicinity of $\epsilon_F$,  since at the node (antinode) the DOS oscillations have a small (large) magnitude. At large temperatures $kT \delta f \gg 1$  the $kT$ interval contains both node (s) and antinode (s) and the averaged square of the DOS oscillations does not depend on the particular location of the beating pattern with respect to $\epsilon_F$.  At this condition  MISO oscillations should be  suppressed. This consideration advocates for a decrease of the MISO amplitude with the temperature in  magnetically entangled subbands.

\subsection{Temperature damping of MISO in magnetically entangled subbands }

We consider 2D electron system with two populated parabolic subbands placed in  a small quantizing perpendicular magnetic field $B_\perp$  and an in-plane magnetic field  $B_\parallel$: ${\bf B}=(B_\perp, B_\parallel)$. In accordance with the presented numerical analysis of the electron spectrum (see also Appendix A) at a non-zero $B_\parallel$ the cyclotron masses, $m_{c1} >m_{c2}$ and frequencies, $\omega_{c1}< \omega_{c2}$, are different. This difference leads to the density of states (DOS) oscillating at different frequencies, $f_i$,  in  different subbands: $f_i=1/\hbar \omega_{ci}$, where index $i$=1(2) corresponds to first (second) subband.   

At a small quantizing magnetic fields  $\omega_{ci} \tau_q < 1$ the main contribution to  MISO comes from the fundamental harmonics of DOS oscillations. The  DOS of $i$-th spatial subband, $\nu_i(\epsilon)$, reads\cite{ando1982,mayer2016a}:    
\begin{equation}
\begin{split}
&\frac{\nu_1(\epsilon\geq 0)}{\nu_{01}}=1-2\delta_1\cos (2\pi f_1 \epsilon ) \\ 
&\frac{\nu_2(\epsilon \geq \ E_g)}{\nu_{02}}=1- 2\delta_2 \cos (2\pi f_2(\epsilon - E_g))
\end{split}
\label{dos}
\end{equation}
where $\nu_{0i}$ represents  DOS at  zero perpendicular magnetic field, $\delta_i=exp(-\pi/\omega_{ci} \tau_q^{(i)})$ is Dingle factor  and $\tau_q^{(i)}$ is the quantum scattering time in  $i$-th subbands. The parameters $\nu_{0i}$ describe DOS in a $kT$ vicinity of the Fermi energy. Within the $kT$ interval the energy dependence of these parameters in a weakly non-parabolic spectrum of 2D electrons, induced by the in the in-plane magnetic field,  is neglected.  
 
The 2D conductivity $\sigma$ is obtained from the following relation:
\begin{equation}
\sigma({\bf B})=\int d\epsilon \sigma(\epsilon)\left(-\frac{\partial f_T}{\partial \epsilon}\right)=\langle \sigma(\epsilon) \rangle
\label{cond}
\end{equation}  
The integral is  an average of the conductivity $\sigma(\epsilon)$ taken essentially for  energies  $\epsilon$  inside the  temperature interval $kT$ near Fermi energy, where $f_T(\epsilon)$ is the electron distribution function at a temperature $T$. \cite{ando1982,ziman} The brackets represent this integral below.  We consider the regime of  high temperatures: $f_i kT \gg 1$. In this regime Shubnikov de Haas oscillations are suppressed but MISO survive.

The conductivity $\sigma (\epsilon)$ is proportional to square of the total density of states: $\sigma (\epsilon) \sim (\nu_1(\epsilon)+\nu_2(\epsilon))^2$.\cite{dmitriev2005,mayer2016a} 
This relation yields the following term leading to  MISO  at small quantizing magnetic fields\cite{raikh1994,raichev2008}:
\begin{equation}
\sigma_{MISO}(\epsilon)=\sigma_D^{(12)}\tilde{\nu_1}(\epsilon)\tilde{\nu_2}(\epsilon)
\label{cond2}
\end{equation}
where $\tilde{\nu_i}(\epsilon)=\nu_i(\epsilon)/\nu_{0i}$ are the normalized density of states in each spatial subband. The parameter  $\sigma_D^{(12)}(B_\perp)$ is Drude like conductivity, accounting  for  inter-subband scattering.\cite{raikh1994,raichev2008}

A substitution of  Eq.(\ref{cond2}) and Eq.(\ref{dos}) into Eq.(\ref{cond}) yields the following expression  for the MISO  of conductivity: 
\begin{equation}
\sigma_{MISO} ({\bf B})=4\sigma_D^{(12)}\delta_1 \delta_2 \langle \cos(2 \pi f_1 \epsilon) \cos(2 \pi f_2 (\epsilon-E_g)) \rangle
\label{cond3}
\end{equation}

An energy integration (see details in Appendix B) yields the final result:
\begin{equation}
\sigma_{MISO} ({\bf B})=2 \sigma_D^{(12)}\delta_1 \delta_2 \frac{X}{\sinh(X)}\cos(2\pi f_2 E_g+2\pi \delta f \epsilon_F)
\label{cond_fin}
\end{equation}
, where parameter $X=2\pi^2kT\delta f$ and $\delta f=f_1-f_2$.  

The obtained expression reproduces the results for disentangled subbands at $B_\parallel$=0T.\cite{raikh1994,raichev2008}  Indeed at $B_\parallel$=0T the difference frequency $\delta f$=0 and  the temperature damping factor $A_{MISO}(T)=X/\sinh(X)=1$. The MISO maxima  correspond to the condition $f_2 E_g=j$, where $j$ is a positive integer,  which is equivalent to Eq.(\ref{miso_eq}) since $f_2=f=1/\hbar \omega_c$ and $E_g=\Delta_{12}$ at  $B_\parallel$=0T.  Finally the MISO magnitude is proportional to the product of two Dingle factors $\delta_1$ and  $\delta_2$.\cite{raikh1994,raichev2008}

For entangled subbands $\delta f>$0 and the temperature damping factor $A_{MISO}(T)=X/\sinh(X)$ decreases the MISO amplitude. This temperature decrease becomes exponential for  $X >$1 since $\sinh(X) \sim \exp(X)$ for $X>$1.   The parameter $X$ is proportional to the temperature and the difference frequency $\delta f=f_1-f_2$. At small in-plane magnetic fields, $B_\parallel$,   the difference frequency is proportional to  $B_\parallel^2$. This is shown in the insert to Figure \ref{spectrum} since  $\delta f=f(m_{c1}-m_{c2})/m_0$ and $(m_{c1}-m_{c2})/m_0 \approx \chi B_\parallel^2$ at small $B_\parallel$, where $\chi$ is a constant . Thus at small in-plane magnetic fields the parameter $X=2\pi^2 kTf \chi B_\parallel^2= [2\pi^2 km_0/(e\hbar)] \chi \tan^2(\theta)TB_\perp$ is proportional to $T$ and $B_\perp$.   At  larger $B_\parallel$ the mass divergence  becomes weaker than $B_\parallel^2$, indicating a presence of  high order terms of $B_\parallel^2$.   Within the order of $B_\parallel^6$ the parameter  $X$ reads:
\begin{equation}
X=\frac{2\pi^2 k m_0}{e\hbar}\chi (1-\xi B_\parallel^2+\eta B_\parallel^4)\tan^2(\theta)TB_\perp
\label{xpar}
\end{equation} 
where $\chi$, $\xi$ and $\eta$ are  constants. In Appendix A the constants  $\chi=1.12\cdot10^{-5}[d(nm)]^2$ and $\xi=1.91\cdot10^{-5}[d(nm)]^2$ are computed analytically  for the magnetically entangled subbands. Below we use the relation (\ref{xpar}) to compare experiments with the expression (\ref{cond_fin}).   

In many respects  the MISO temperature damping factor $A_{MISO}(T)$ is similar  the one for  Shubnikov de  Haas oscillations, $A_{SdH}(T)=X_{SdH}/\sinh(X_{SdH})$, where $X_{SdH}=2\pi^2 kT/(\hbar \omega_{ci})$.\cite{shoenberg1984} The main difference is that the factor $A_{MISO}$ depends on the  difference frequency $\delta f$ whereas the $A_{SdH}$ depends on the frequency $f_i=1/\hbar \omega_{ci}$. For parabolic subbands with the same  masses $\delta f$=0 and the MISO damping factor $A_{MISO}$=1 is irrelevant.  The  MISO damping factor is important for non-parabolic spectra or parabolic spectra  with different cyclotron masses in two subbands.

\section{Temperature dependence of MISO in tilted magnetic field} 
\begin{figure*}[t]
\vskip -0 cm
\includegraphics[width=15cm]{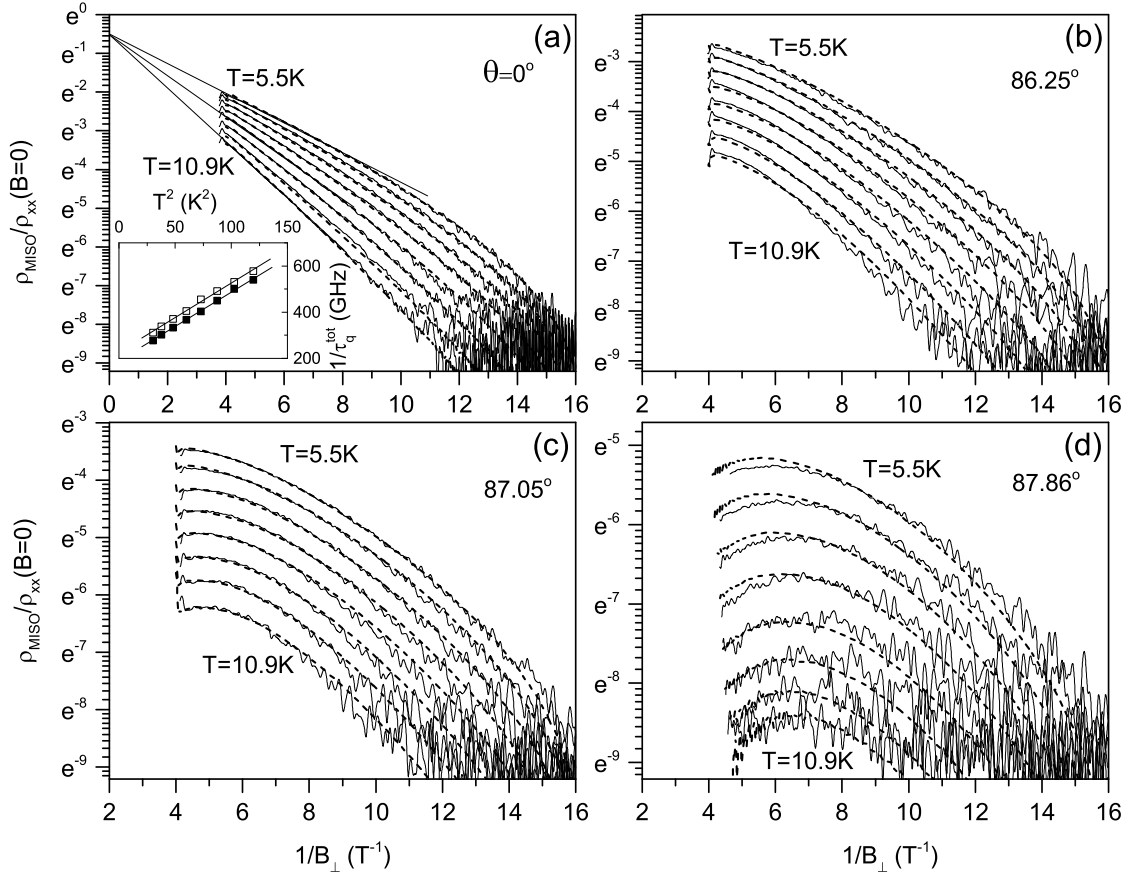}
\caption{Dependence of normalized MISO amplitude $\rho_{MISO}/\rho_{xx}(0)$ on reciprocal magnetic field, $1/B_\perp$, at different temperatures from top to bottom T=5.5, 6.14, 6.93, 7.74, 8.54, 9.34, 10.13 and 10.93K. and at  angles as labeled.  Solid lines represent experimental data. Dashed lines are numerical computations of MISO magnitude multiplied with normalizing function $F_N(B_\perp)=0.55\cos(0.096/B_\perp)$.   (a) The numerical computations use quantum scattering times $\tau_q^{(1)}=\tau_q^{(2)}$ as fitting parameter to match with the experiment at different temperatures. In the insert, filled symbols present the obtained   total quantum scattering rate: $1/\tau_q^{tot}=2/\tau_q^{(1)}$. Open symbols present the rate determined from slopes of the thin straight lines shown in (a) ; (b)-(d) determined in (a)  rates $1/\tau_q^{tot}$ are used to compute MISO magnitude. The computed dependencies are shifted vertically  to match with experiment, using normalizing factor $K(T)$. $d=26$nm. Sample  S1. }
\label{ang}
\end{figure*} 
In this section we compare the described  model  above and numerical computations of MISO  with experiment. We start with the comparison between the numerical estimations and experiment. 

Figure  \ref{ang} presents dependence of MISO amplitude on reciprocal magnetic field, $1/B_\perp$, measured at different temperatures between 5.5K and 10.9K. Panels (a)-(d) show the dependencies taken at different angles $\theta$ between the normal to 2D layer and the direction of the magnetic field $\bf B$. The dashed lines present results of numerical computations of MISO magnitude.

Figure  \ref{ang}(a) presents the dependencies taken at  $\theta=0^0$. At this angle the entanglement between subbands is absent and $A_{MISO}=$1. The MISO magnitude decreases  strongly with the reciprocal magnetic field, $1/B_\perp$. This decrease is due to the exponential decrease of Dingle factors $\delta_i$  with $1/B_\perp$: $\delta_i=exp(-\pi/\omega_{ci}\tau_q^{(i)})$. In accordance with Eq.(\ref{cond_fin}),  the MISO magnitude is proportional to  the product of the Dingle factors.  For disentangled subbands the cyclotron frequencies $\omega_{c1}$ and $\omega_{c2}$ are the same since $m_{c1}=m_{c2}=m_0$. Thus, the dependencies of the MISO amplitude on $1/B_\perp$, plotted in semi-log scale, should be straight lines with the slope  proportional to  the sum of quantum scattering rates in two subbands: $1/\tau_q^{(1)}+1/\tau_q^{(2)}$.  In  Figure  \ref{ang}(a) thin solid straight lines present  the linear approximation of the measured dependencies. At higher temperature the slope of the lines becomes larger, indicating an increase of the quantum scattering rate with the temperature increase.  In the insert to  Figure  \ref{ang}(a) open symbols presents  the temperature dependence of  total quantum scattering rate $1/\tau_q^{tot}=1/\tau_q^{(1)}+1/\tau_q^{(2)}$, extracted from these slopes.  

 A noticeable feature of the linear  approximation is the convergence of the straight lines to the  single point at $1/B_\perp$=0T. This feature follows from Eq.(\ref{cond_fin}) since   $\delta_1\delta_2 \rightarrow$1 and, thus,  becomes temperature independent at $1/B_\perp \rightarrow$0. Another noticeable feature is the apparent deviation of the measured dependencies from the straight lines at $1/B_\perp >$10 (1/T). The origin of this deviation is  under investigation and is not the  focus of this study. We have found that a normalization of  Eq.(\ref{cond_fin}) by a temperature independent function $F_N(B_\perp)$ leads to a good agreement between experiment and the model.    

In Figure \ref{ang}(a) the dashed lines present results of the numerical evaluation of the MISO magnitude. For each temperature the MISO magnitude is evaluated numerically with only one fitting parameter - the total quantum scattering rate $1/\tau_q^{tot}$.   
The computed dependence is multiplied by the normalizing function $F_N(B_\perp)=0.55\cos(0.096/B_\perp)$, which bends down the linear dependence at $1/B_\perp >$10 providing  good agreement with the experiment. Obtained via this procedure the  total scattering rate is shown by filled symbols in the insert to the figure. This scattering rate is found to be slightly lower than the one obtained via the first procedure (open symbols). Both dependencies demonstrate essentially the same variations of the quantum scattering rate with the temperature: $\delta (1/\tau_q^{tot}) \sim T^2$, indicating the dominant contribution of the electron-electron scattering to the quantum electron lifetime.\cite{mamani2008,goran2009,dietrich2012} 

 For entangled subbands the cyclotron frequencies $\omega_{c1}$ and $\omega_{c2}$ are  different  since $m_{c1}>m_{c2}$. The difference leads to variations of the product of Dingle factors with the in-plane magnetic field in Eq.(\ref{cond_fin}). Both numerical and analytical investigations of these variations demonstrates  weak (within few percents ) corrections to MISO magnitude in the studied range of parameters.  At $\tau_q^{(1)}=\tau_q^{(2)}$       these corrections are absent. Below we neglect these corrections and use $\tau_q^{(1)}=\tau_q^{(2)}$.       

\begin{figure}[t]
\vskip -0 cm
\includegraphics[width=\columnwidth]{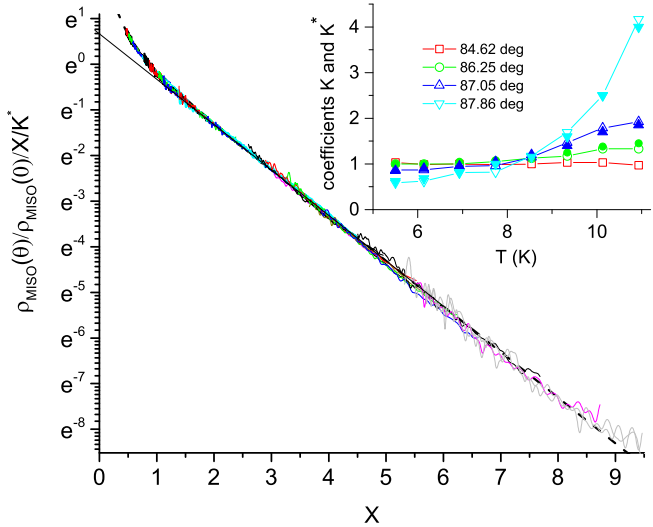}
\caption{Dependence of the ratio $R$ of MISO magnitude obtained at angle $\theta$ to the one at  $\theta$=0 deg normalized by $X$ and $K^*$: $R^*=R/X/K^*$  on parameter $X$ at different temperatures $T$: 5.5, 6.14, 6.93, 7.74, 8.54, 9.34, 10.13 and 10.9K and different angles $\theta$=84.62, 86.25, 87.05 and 87.86 deg (see text for detail).   The parameter $X$ is computed from Eq.(\ref{xpar}), using $\chi=1.12\cdot 10^{-5}[d(nm)]^2$, $\xi=1.91\cdot 10^{-5}[d(nm)]^2$, which are evaluated in Appendix A, and $\eta=4\cdot 10^{-10}[d(nm)]^4$ at $d$=26nm. Dashed line presents the dependence $1/\sinh(X)$ expected from Eq.(\ref{cond_fin}). Thin straight line presents the linear dependence $R/X$ vs $X$ with a unity slope and intersect $y_0$=2, expected from Eq.(\ref{cond_fin}) at $X>$1. The insert presents temperature dependence of normalizing coefficients $K$(filled symbols) and $K^*$(open symbols)  at different angles as labeled.      }
\label{ancom}
\end{figure}

Figure \ref{ang}(b) presents the magnetic field dependence of the MISO magnitude at $\theta=86.25^0$. At this angle the magnetic entanglement between two subbands leads to modifications of the MISO magnitude.  Indeed at $1/B_\perp \approx$5 (1/T) and $T$=5.5K the relative MISO magnitude is 0.058, which  is considerably smaller the one shown in  panel (a) - 0.094. At higher temperature T=10.9K the ratio between these two magnitudes becomes even smaller: 0.37.   The numerical evaluations  demonstrate the decrease of the MISO magnitude with the magnetic field tilt  and the temperature and mostly capture  the changes in the dependence shape.   To   better compare variations of the shape of the dependencies  the overall magnitude of the numerical MISO is  multiplied  by factor $K(T)$, which is shown in the insert to Fig. \ref{ancom}. In Fig.\ref{ang}(b)-(d) the factor $K$ moves the computed dependencies vertically providing a better overlap with the experiment.   

Figures \ref{ang}(c) and (d) present the magnetic field dependence of the MISO magnitude at $\theta=87.05^0$ and $\theta=87.86^0$.  At  larger tilts the entanglement between subbands becomes stronger leading to  stronger suppression of the MISO magnitude. The numerical computations continue to demonstrate good correlations with  the shape of the magnetic field dependencies at different temperatures. These dependencies are not only quantitatively but qualitatively different  from the ones shown in panel (a) for the disentangled subbands. In particular the convergence of the responses  at $1/B_\perp \rightarrow$0, which is apparent in panel (a), disappears in panels (c) and (d). Another noticeable feature  is a consistent increase of variations of the normalizing coefficient $K$ with the temperature and the tilt, which is shown in the insert to Fig. \ref{ancom}.  This MISO property will be  discussed later.

All numerical dependencies,  shown in panels (b), (c) and (d),  are obtained at fixed $d$=26 nm, providing the best agreement  with the shapes of experimental dependencies.  The quantum scattering rates are determined from the  response of  disentangled subbands shown in panel (a). Thus, in the panels (b)-(d) the only variable  fitting parameter,  is the normalizing factor $K$, which moves the dependencies vertically but does not change their shape. Thus,  as for the functional dependence the  presented  in panels (b)-(d) comparison between experiment and the model uses only one fitting parameter - the width of the quantum well $d$.  The obtained width $d$=26 nm coincides with the actual width of the studied 2D layer. Thus the presented model  captures  the variations of the shape of the dependency of MISO on  $1/B_\perp$.  

Presented in Fig.\ref{ang} comparison with the numerical MISO is done under  assumptions  that the quantum scattering rates $1/\tau_q^{(i)}$  and the Drude like conductivity $\sigma_D^{(12)}$ do not vary  with the entanglement between subbands. The obtained agreement supports these assumptions, which we follow below.

To reveal  the temperature damping factor $ A_{MISO}(X)=X/\sinh(X)$ we compare our experimental data with the analytical expression (\ref{cond_fin}) containing this factor. 
There are  other factors ($\delta_i , \sigma^{(12)}_{D}$) entering the expression. The presented  comparison above with the numerical MISO  as well as analytical considerations indicate that  the product of these factors  vary very  weakly  with the entanglement between subbands. Below we neglect these variations.  To remove effects of these factors in the comparison between Eq.(\ref{cond_fin}) and experiment,  we divide each dependence  in panels (b)-(d) (entangled subbands) by the dependence from panel (a) (disentangled subbands)  taken at the same temperature $T$. This ratio  $R_{exp}=\rho_{MISO}(\theta)/\rho_{MISO}(0)$ is compared with the one obtained from Eq.(\ref{cond_fin}).  In accordance with Eq.(\ref{cond_fin}) at $\tau_q^{(1)}=\tau_q^{(2)}$ the ratio of the MISO magnitudes $R_{mod}=X/\sinh(X)$ and depends only on the parameter $X$. Thus, plotted vs $X$,  the ratio $R_{exp}(X)$ should  follow $A_{MISO}(X)=X/\sinh(X)$.  To facilitate the comparison at $X >$1 both ratios are divided by $X$, yielding $R_{mod}/X \approx 2\exp(-X)$  at $X >$1. At large X  $\ln(R/X)$ vs $X$ is, thus, straight line with a unity slope  intersecting  $y$-axis at $y_0$=2.  

\begin{figure}[t]
\vskip -0 cm
\includegraphics[width=\columnwidth]{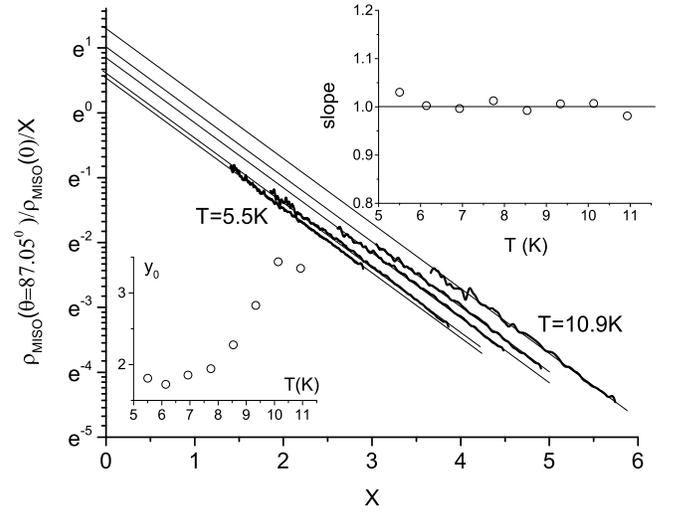}
\caption{Dependence of  ratio of MISO magnitude at $\theta=87.05^0$ to the one at  $\theta=0^0$, normalized by $X$, on parameter $X$ at different temperatures $T$: 5.5, 7.74, 8.54, 9.34 and 10.9K.   The parameter $X$ is computed from Eq.(\ref{xpar}), using $\chi=1.12\cdot 10^{-5}[d(nm)]^2$, $\xi=1.91\cdot 10^{-5}[d(nm)]^2$, (see Appendix A), and $\eta=4\cdot 10^{-10}[d(nm)]^4$. Thin straight lines present linear dependencies with a unity slope, expected from Eq.(\ref{cond_fin}). Upper insert presents temperature variations  of  slope magnitude, obtained from  linear fit  of the normalized ratio. Lower insert presents temperature evolution of the  intersect $y_0$ of the linear fit   with $y$-axis.   }
\label{ang87}
\end{figure}

Figure \ref{ang87}  presents the  dependence of the ratio $R_{exp}/X=\rho_{MISO}(\theta)/\rho_{MISO}(0)/X$ on the parameter $X$ for  data at $\theta=87.05^0$.    The parameter $X$ is evaluated from Eq.(\ref{xpar}), using parameters $\chi=1.12\cdot 10^{-5}[d(nm)]^2$ and  $\xi=1.91\cdot 10^{-5}[d(nm)]^2$, computed  in Appendix A   and parameter $\eta=\eta_0[d(nm)]^4$, where $\eta_0$ is a fitting parameter. At $d$=26 nm for all  temperatures the experimental dependencies $\ln(R/X)$ vs $X$  follow  the straight lines with  unity slope. Some of the straight  lines and the dependencies are shown  in Figure \ref{ang87} . The upper insert to Figure \ref{ang87} demonstrates the magnitude of slopes  obtained by a linear fit of the data. The slope magnitudes fluctuate around the expected value 1.  

At $T$=5.5K the intersect of the corresponding straight line with y-axis yields $y_0 \approx$1.72. This value is slightly below the expected value 2. With an increase of the temperature the intersect $y_0$ increases. The lower insert presents the  increase of the intersect $y_0$ with the temperature obtained from the linear fit of the data.  Thus, similar to the comparison with the numerical  MISO, shown in Figure \ref{ang},  the comparison  in Figure \ref{ang87}  advocates for an additional factor $K^*(T,\theta)$ controlling the MISO magnitude. 

At different temperatures and angles the normalizing factor $K^*$ is determined by the best overlap of experimental data with the expected dependence $1/\sinh(X)$. To cancel effects related to this factor the experimental data $R_{exp}=\rho_{MISO}(\theta)/\rho_{MISO}(0)$  is divided by  $K^*(T, \theta)$.     This procedure leads to a collapse of  experimental dependencies on the single curve $1/\sinh(X)$, shown in Fig.\ref{ancom}.

Figure \ref{ancom} presents dependence of the normalized ratio $R^*=\rho_{MISO}(\theta)/\rho_{MISO}(0)/X/K^*$ on the parameter X for different temperatures and angles. The figure shows that for a broad range of temperatures and subband entanglement the normalized MISO magnitude, $R^*$, depends on  the single parameter $X$,   demonstrating good agreement with the modified MISO temperature damping factor $A_{MISO}/X=1/\sinh(X)$, shown by the dashed line in the figure. Thus, both  comparisons, which are presented in Figure \ref{ang} and Figures \ref{ang87} and \ref{ancom},  indicate that  variations of MISO magnitude with  the reciprocal magnetic field $1/B_\perp$, temperature $T$, angles $\theta$  agree with the model and are controlled by MISO temperature damping factor  $A_{MISO}=X/\sinh(X)$. 

Both comparisons indicate also that there is another controlling factor  $K^*(\theta, T) \approx K(\theta, T)$, which is beyond the presented model.  The insert to Fig.\ref{ancom} shows temperature dependencies  of normalizing coefficients $K$ (filled symbols) and $K^*$ (open symbols), obtained by different fitting procedures. Both procedures indicate  the same temperature increase of both   factors at a given angle.  The data shows that the temperature variations of the parameters $K$ and $K^*$ are larger at larger  $\theta$. 

At   large angles $\theta=87.05^0$ and $\theta=87.86^0$ the unity slope of the dependencies $R^*(X)$ is observed for all temperatures. However at  smaller angles ($\theta=84.62^0$ and $\theta=86.25^0$) and high temperatures ($T>$9K) the dependencies $R^*(X)$ demonstrate slopes with magnitudes which are distinctly  smaller than the unity.  These dependencies are not shown in Fig.\ref{ancom}. The presence of these deviations suggests a transitional function $F_{tr}(\delta f, \theta,T)$ between  regimes of a weak and strong  subband entanglement with a property $F_{tr}(\delta f, \theta,T)\rightarrow K^*(\theta, T)$ at a large $X$.  The transitional function  has not been investigated in this study.   At  large angles $\theta$ and  temperatures (large $X$), where  the normalizing coefficient $K^*$ and the function $F_{tr}(\delta f, \theta, T)$  are measurable,  the access to small $X$ requires a very small $B_\perp$ (see Eq.(\ref{xpar}). At this small $B_\perp$ the Dingle factors   strongly suppress  the MISO amplitude making the amplitude measurements  not accurate. Measurements at smaller angles indicate  the presence of  the transitional function. However the magnitude of this function is  small, making an analysis of the function to be not informative.   

\subsection{Effects of electron-electron interaction on MISO}

Both Figure \ref{ang87} and the insert to Figure \ref{ancom} demonstrate an increase  of the deviation between the experiment and  model  with the temperature increase. 
The increase  of the deviation  correlates with the increase of the temperature dependent  contribution to the electron lifetime. Indeed  the insert to Figure \ref{ang}(a) shows that at $T$=10.9 K the contribution of electron-electron scattering to the quantum scattering rate  is about 4 times larger  than at $T$=5.5K and becomes dominant. This correlation suggests that effects of electron-electron interaction or Fermi-liquid effects may play important role leading to the deviation between Eq.(\ref{cond_fin}) and experiment.  Indeed, although   ignored in the presented model, such effects  are important for quantum oscillations, resulting  in a renormalization of the  electron mass and g-factor - the effects, which have been intensively investigated both theoretically and experimentally for several decades\cite{ando1982}. 

Effects of the electron-electron interactions on the quantum scattering time, controlling the magnitude of quantum oscillations, are  less frequently studied.   Existing theory predicts  that the amplitude of the fundamental harmonic of  SdH oscillations is resilient to the temperature variations of the quantum scattering time, induced by the electron -electron interaction.\cite{maslov2003,mirlin2006} In other words the quantum scattering time, entering the Dingle factor for the fundamental harmonic of SdH oscillations, is a temperature independent parameter. This can be considered as a result of  a modification of the electron lifetime by the electron-electron interaction. The modification leads to contributions, enhancing the SdH amplitude and  compensating  the temperature dependent part of the quantum scattering rate in the Dingle  factor.   In contrast  the quantum scattering rate, entering the Dingle factor for the MISO amplitude, is temperature dependent property as shown in the insert to Figure \ref{ang}(a).  

To the best of our knowledge  Fermi liquid  effects related  to MISO in magnetically entangled subbands have not been  investigated. Assuming a similarity of the Fermi liquid contributions  to the magnitude of SdH oscillations and MISO in entangled subbands, one should expect a relative increase of the MISO magnitude, which may explain the increase of the factors $K$ and $K^*$ with the temperature. The resilience of SdH amplitude to the electron-electron interactions can be obtained via an account of  the interaction induced dependence of the electron-electron scattering rate on the energy $\epsilon$.\cite{dmitriev2009}   The electron-electron collision rate for an electron at energy $\epsilon$ counted from the Fermi energy $\epsilon_F$ is 

\begin{figure}[t]
\vskip -0 cm
\includegraphics[width=\columnwidth]{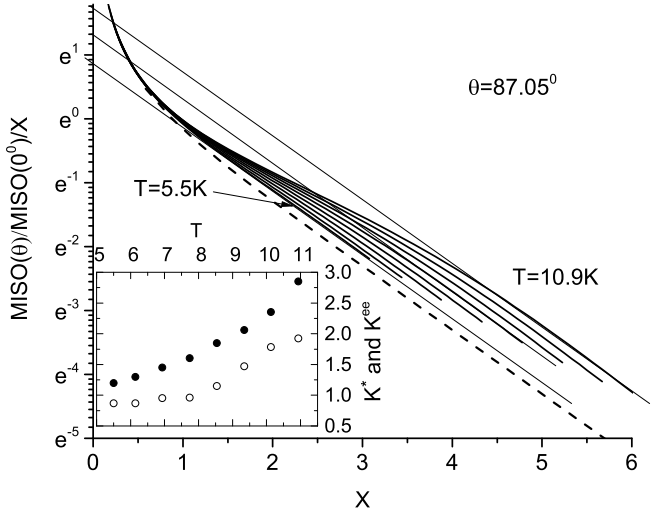}
\caption{Dependence of  ratio of MISO magnitude at $\theta=87.05^0$ to the one at  $\theta=0^0$, normalized by $X$ on parameter $X$. The dependence is computed  at $\epsilon_F^*=$8 meV and  different temperatures from bottom to top $T$=5.5, 6.14, 6.93, 7.74, 8.54, 9.34, 10.13 and 10.9K using Eq.(\ref{ratio}).   The parameter $X$ is computed from Eq.(\ref{xpar}), using $\chi=1.12\cdot 10^{-5}[d(nm)]^2$, $\xi=1.91\cdot 10^{-5}[d(nm)]^2$, obtained in Appendix A, and $\eta=4\cdot 10^{-10}[d(nm)]^4$. Thin straight lines present linear dependencies with a unity slope. Dashed line displays free electron response $1/\sinh(X)$. The  insert shows temperature evolution of factors $K^*$ and $K^{ee}$, characterizing maximal deviation of the experimental and model data from the free electron response.}
\label{miso_ee}
\end{figure}

\begin{equation}
\frac{1}{\tau_{ee}(\epsilon, T)}=\frac{\epsilon^2+\pi^2(kT)^2}{4\pi \hbar \epsilon_F}\ln\frac{q_s v_F}{\max (kT, \hbar \omega_c(\omega_c\tau_{tr})^{1/2})  }
\label{t_ee}
\end{equation} 
, where $v_F$ is Fermi velocity, $\tau_{tr}$ is transport scattering time and $q_s=2\pi e^2\nu$ is inversion screening length.\cite{dmitriev2005,dmitriev2009}  

The energy dependence of the electron scattering rate makes  the Dingle factors $\delta_i$ to be energy dependent parameters:
\begin{equation}
\delta_i(\epsilon, T)=\exp\Big(-\frac{\tau_{im}^{-1}+\tau_{ee}^{-1}(\epsilon,T) }{\omega_{ci}/\pi}\Big)
\label{d_ee}
\end{equation}  
, where $\tau_{im}$ is quantum scattering time due to impurity scattering. The time $\tau_{im}$ does not depend  on the temperature while the electron-electron scattering time $\tau_{ee}$ is temperature dependent. The time $\tau_{ee}$ provides the $T^2$ contribution to the quantum scattering rate shown in the insert to Fig. \ref{ang}(a) for the disentangled subbands. 

The energy dependence of the Dingle factors $\delta_i$ is not accounted in the presented above analysis. The effect of the energy dependence of the $e-e$ scattering rate on  the relative MISO magnitude: $\rho_{MISO}(\theta)/\rho_{MISO}(0^0)=\sigma_{MISO}(\theta)/\sigma_{MISO}(0^0)$ is evaluated below.   Substitution of the relations (\ref{cond2}), (\ref{dos}) and (\ref{d_ee}) into Eq.(\ref{cond}) leads to the following expression for the relative MISO magnitude:  

\begin{equation}
\frac{\rho_{MISO}(\theta)}{\rho_{MISO}(0^0)}=\frac{\langle \exp(-\epsilon^2/\epsilon_0^2) \cos(2 \pi \delta f \epsilon) \rangle}{\langle \exp(-\epsilon^2/\epsilon_0^2) \rangle}
\label{ratio}
\end{equation}       
, where $\epsilon_0=(2\epsilon_F^* \hbar \omega_c)^{1/2}$.  In the estimation  a possible difference in the $e-e$ scattering rate in two subbands and the temperature/magnetic field dependencies of the logarithmic  factor in Eq.(\ref{t_ee}) are ignored. As a result  in Eq.(\ref{ratio}) the only fitting parameter  is $\epsilon_F^* \sim \epsilon_F^{(i)}/ \ln(q_sv_F^{(i)}/\max(kT, \hbar\omega_c(\omega_c)^{1/2} $. 

Figure \ref{miso_ee} demonstrates the dependence of normalized  relative MISO magnitude, $\rho_{MISO}(\theta)/\rho_{MISO}(0^0)/X$ on parameter X obtained from Eq.(\ref{ratio}) at angle $\theta$=87.05$^0$,  temperatures $T$= 5.5, 6.14, 6.93, 7.74, 8.54, 9.34, 10.13 and 10.9K and $\epsilon_F^*$=8 meV. The angle and temperatures corresponds to the experimental dependencies of the normalized relative MISO magnitude presented in Fig.\ref{ang87}. In Fig.\ref{miso_ee} the dashed line shows the dependence $1/\sinh(X)$ for free 2D electrons computed at $\epsilon_0 \rightarrow \infty $.  The obtained  behavior suggests that the relative MISO magnitude can be presented as  a product of  $X/\sinh(X)$ and a finite function $F_{tr}(X,\theta, T)$: 

\begin{equation}
\frac{\rho_{MISO}(\theta)}{\rho_{MISO}(0^0)}=F_{tr}(X,\theta, T) \frac{X}{\sinh(X)}
\label{ratio2}
\end{equation}   

Below we investigate  properties of the function $F_{tr}(X,\theta, T)$. 
In Fig.\ref{miso_ee} at small $X<$1  the dependencies converge for all temperatures. This is related to the reduction of the difference frequency:  $\delta f  \rightarrow 0$ at $X  \rightarrow 0$  since $\delta f$ is proportional to $X$. At $\delta f   \rightarrow 0$ in Eq.(\ref{ratio}) the cosine function tends to 1 and the ratio of the two integrals approaches  unity.  Thus at $X  \rightarrow 0$ the function $F_{tr}(X,\theta, T)  \rightarrow 1$ since $X/\sinh(X) \rightarrow 1$.

 At  large $X  \rightarrow \infty$ but a finite temperature  the function $F_{tr}(X,\theta, T)$    also tend to unity.  To understand this property we note that in accordance with  Eq.(\ref{xpar}) a large $X$ corresponds to a large $B_\perp$ and, thus,  to  large  $\hbar \omega_c $ and $\epsilon_0$. At $\epsilon_0 \gg kT$ in Eq.(\ref{ratio}) the Gaussian functions can be neglected that leads to the free electron result (\ref{cond_fin}).
 
 At an intermediate X the function  $F_{tr}(X,\theta, T)$ deviates from unity and reaches a maximum. The increase  of the function $F_{tr}(X,\theta, T)$ from the unity  is a result of the electron-electron interaction and, thus, is a Fermi liquid effect. The electron-electron interaction leads to a decrease of the quantum lifetime of quasiparticles with the energy $\epsilon$ away from the Fermi energy.\cite{maslov2003,mirlin2006}  Eq.(\ref{t_ee}) and Eq.(\ref{d_ee}) take into account this lifetime decrease  and yields in Eq.(\ref{ratio}) the Gaussian $\exp(-\epsilon^2/\epsilon_0^2)$, which enhances the MISO amplitude.     Mathematically the effect is due to  a reduction of  the range of the energy integration  in Eq.(\ref{ratio})  from $(-kT, kT)$, settled by the distribution function $f_T$ for free electrons,  to a smaller range, which for the interacting electrons is additionally affected  by the range narrowing factor $\exp(-\epsilon^2/\epsilon_0^2)$.    The energy averaging of  the oscillating content ($\cos(2 \pi \delta f \epsilon)$) in  narrower energy interval leads to a suppression of the averaging and results in a larger value of the integral and, thus, the function $F_{tr}(X,\theta, T)$. \cite{dmitriev2009}
 
In the experimentally studied range of parameters, the maximum of the function $F_{tr}(X,\theta, T)$ appears to be  quite flat and can be approximated by a straight horizontal line, which acquires unity slope in Fig.\ref{miso_ee}. This property agrees with the experiment. Three of these lines are shown in  Fig.\ref{miso_ee}.  A coefficient $K^{ee}(T) \approx \max[F_{tr}(X,T)]$ characterizes the vertical displacement of these lines  from the free electron response $1/\sinh(X)$ (dashed line).    Figure \ref{miso_ee} demonstrates that the coefficient  $K^{ee}(T)$ increases with the  temperature. This behavior is also in agreement with the experiment shown in Fig.\ref{ang87}. 
 
 The insert to Fig.\ref{miso_ee} demonstrates a comparison between  coefficient $K^*$, obtained from experimental data presented in Fig.\ref{ang87} and coefficient $K^{ee}$, obtained from the model data presented in Fig.\ref{miso_ee}.  At $\epsilon_F^*$=8 meV both coefficients $K^*$, $K^{ee}$ and variations of these coefficients with the temperature are close to each other. Furthermore an evaluation of the temperature dependence of the quantum scattering rate, using the temperature dependent part of Eq.(\ref{t_ee}),  yields $\tau_q^{-1}(T)-\tau_q^{-1}(0K)=\tau_{ee}^{-1}(\epsilon=0)=\pi(kT)^2/(4 \hbar \epsilon_F^*)\approx$1.2(GHz)$T^2$.  This value  is close to the inelastic scattering rate obtained in the experiment at $\theta$=0$^0$ and shown in the insert to Fig.\ref{ang}(a): $\tau_q^{-1}(T)-\tau_q^{-1}(0K) \approx $1.5(GHz)$T^2$.   Thus, the account of the  electron-electron interaction   improves   the  agreement  between the experiment and  model, revealing the interaction  induced enhancement of MISO amplitude.

\section{Summary}

Magneto-intersubbands resistance oscillations (MISO)  of highly mobile 2D electrons  in  symmetric GaAs  quantum wells with two populated subbands  are studied at different temperatures and at different angles $\theta$ between  magnetic field ${\bf B}$  and the normal to 2D layer. The experiments indicate that the MISO magnitude decreases strongly with the  temperature. For  angles $\theta<80^0$  the MISO reduction is  related to the increase of the quantum scattering rate due to the enhancement  of electron-electron scattering  at high temperatures. For  angles  $\theta> 80^0$   new regime of strong MISO damping with the temperature is identified. 

Proposed model considers the  magnetic entanglement between subbands, which is  induced by in-plane magnetic field, as the main reason of the new temperature damping. The entanglement changes the electron spectrum and leads to different cyclotron masses in two subbands. As a result the density of states exhibits  beating with the  difference frequency $\delta f$ proportional to the mass difference. The model yields  universal  temperature damping factor $A_{MISO}=X/\sinh(X)$, where $X=2\pi^2 kT \delta f$. 

A comparison of the model with the experiment  demonstrates the presence of the factor $A_{MISO}$ but  indicates an additional factor $K(T)$, which is beyond the free electron model. The factor $K$ leads to an effective enhancement of the MISO amplitude at high temperatures.  An account of the electron-electron interaction  explains the enhancement of the MISO amplitude and reveals the Fermi liquid origin of the factor K.

This work was supported by the National Science Foundation (Division of Material Research - 1702594) and  by the Russian Foundation for Basic Research (project no. 20-02-00309).

\section{Appendix A}
In this section the spectrum of the entangled subbands is computed at $\theta$=90 deg. The cyclotron masses, $m_{ci}$ and difference frequency $\delta f \sim (m_{c1}-m_{c2})$,  are evaluated then  for the quasiclassical  electron motion in a small $B_\perp$.   The goal is  estimation of the variations of the parameter $X \sim \delta f$ with the magnetic field $B_\parallel$  leading to Eq.(\ref{xpar}). 

At $B_\perp$=0 T ($ {\bf B}=(-B_\parallel, 0, 0)$) the Hamiltonian (\ref{ham}) is presented  in the following form:
\begin{equation}
\begin{split}
&H=\frac{\hbar^2 k_x^2}{2m_0}+\frac{\hbar^2 (k_y+\frac{eB_\parallel z}{\hbar})^2}{2m_0}+\frac{\hbar^2 k_z^2}{2m_0}+V(z)= H_0+H_1\\
&H_0=\frac{\hbar^2 }{2m_0}( k_x^2+k_y^2 +k_z^2)+V(z)\\
&H_1=\hbar\omega_\parallel k_y z+\frac{1}{2}m_0\omega_\parallel^2 z^2
\end{split} 
\label{ham2}
\end{equation}
,where $\omega_\parallel =eB_\parallel/m_0$ is the cyclotron frequency in in-plane magnetic field, $B_\parallel$.  At $B_\parallel$=0T the corresponding eigenfunctions  $\vert {\bf k},\xi\rangle$ of the system  are plane waves, propagating in $x-y$ plane, and standing waves in $z$-direction  , where wave vector ${\bf k}$  describes the  lateral motion and $\xi=S, AS$ describes the symmetric (S) and antisymmetric (AS) configurations of the wave function in the $z$-direction (vertical quantization):  $\vert {\bf k},S\rangle=\vert {\bf k}\rangle (2/d)^{1/2}\cos(\pi z/d)$ and  $\vert {\bf k},AS\rangle=\vert {\bf k}\rangle (2/d)^{1/2}\sin(2\pi z/d)$. 

Using  functions $\vert {\bf k},\xi\rangle$ as the basis set , one can present the Hamiltonian as a 2$\times$2 matrix:    
\begin{equation}
\begin{split}
&H_{ij}=\epsilon_i^0 \delta_{ij}+(1-\delta_{ij}) h_{12}\\
&\epsilon_i^0=\frac{\hbar^2}{2m_0}( k_x^2+k_y^2)+E_i+\frac{1}{2}m\omega_\parallel^2 Z_i^2\\
&h_{12}=\hbar\omega_\parallel k_y Z_0
\end{split}
\label{ham3}
\end{equation}
, where $\delta_{ij}$ presents 2$\times$2 unit matrix, $Z_0=16d/(9\pi^2)$, $Z_1^2=(1/12-1/(2\pi^2))d^2$ and $Z_2^2=(1/12-1/(8\pi^2))d^2$.  Indexes $i,j$=1,2 describes first (1) and second (2) subbands.  Energy $E_i$ corresponds to the bottom of $i$-th subband at $B_\parallel$=0T.  

At $h_{12} \ll \epsilon_2^0-\epsilon_1^0$ diagonalization of the Hamiltonian $H_{ij}$ leads to the following spectrum: 
\begin{equation}
\begin{split}
&\epsilon_i({\bf k})\approx \epsilon_i^0({\bf k})  \pm \frac{h_{12}^2}{\epsilon_2^0-\epsilon_1^0}(1-\frac{h_{12}^2}{(\epsilon_2^0-\epsilon_1^0)^2}) \approx \\
& \approx \epsilon_i^0({\bf k})  \pm A \frac{\hbar^2k_y^2}{2m_0}\Big(1-\frac{A}{E_g} \frac{\hbar^2k_y^2}{2m_0}\Big)=\\
&=E_i+\frac{\hbar^2k_x^2}{2m_{xi}}+\frac{\hbar^2k_y^2}{2m_{yi}} \mp \gamma_0 k_y^4\\
&A=\frac{2m_0Z_0^2}{E_g}\omega_\parallel^2; E_g=\Delta_{12}+\frac{m_0(Z_2^2-Z_1^2)}{2}\omega_\parallel^2
\end{split}
\label{spec2}
\end{equation}
, where lower (upper) sign corresponds to the first ($i$=1) (second (i=2)) subband, $\gamma_0=\hbar^4A^2/[(2m_0)^2E_g]$ and $\Delta_{12}=E_2-E_1$. Eq.(\ref{spec2}) indicates that due to the presence of the in-plane magnetic field the spectrum is  anisotropic but still parabolic in the lowest order of $B_\parallel$ ($\sim B_\parallel^2)$. The parameter  $A$ controls the strength of the anisotropy leading to an increase (decrease)  of the mass, $m_{y1}=m_0/(1-A)^{1/2}$ ($m_{y2}=m_0/(1+A)^{1/2}$)  in $y$-direction  for lower (upper) subband. In $x$-direction masses  do not change: $m_{xi}=m_0$.

For a parabolic spectrum the cyclotron mass is $m_c=(m_xm_y)^{1/2}$. \cite{ziman} To compute the cyclotron masses in the vicinity of Fermi energy $\epsilon_F$ for the non-parabolic spectrum  we use the relation $m_c=(\hbar^2/2\pi)(\partial S/\partial \epsilon)$, where $S(\epsilon)$ is the area within the contour $\epsilon_i({\bf k})=\epsilon_F$.\cite{ziman} For the spectrum (\ref{spec2}) the result is
\begin{equation}
\begin{split}
&m_{c1}=(m_{xi}m_{y1})^{1/2}\Big(1 - \frac{3}{4}\frac{m_{y1}^2}{m_0^2}\frac{A^2}{E_g}\epsilon_{F1} \Big)\\
&m_{c2}=(m_{xi}m_{y2})^{1/2}\Big(1 + \frac{3}{4}\frac{m_{y2}^2}{m_0^2}\frac{A^2}{E_g}\epsilon_{F2} \Big)
\end{split}
\label{m_c}
\end{equation}
, where $\epsilon_{Fi}$ is Fermi energy counted from the bottom of $i$-th subband. The result agrees with the numerical computation of the cyclotron masses presented in the insert to Fig.\ref{spectrum}: $m_{c1}$ ($m_{c2}$) increases  (decreases) with the in-plane magnetic field. Furthermore the sum of the masses stays the same: $m_{c1}+m_{c2}=2m_0$  within  the computed order $B_\parallel^4$.

Within the same order  for difference frequency Eq.(\ref{m_c}) yields:

\begin{equation}
\begin{split}
&\delta f \approx f\Big( A -\frac{3}{4}\frac{\epsilon_{F1}+\epsilon_{F2}}{E_g} A^2 \Big)\approx  f \Big[\chi B_\parallel^2(1-\xi B_\parallel^2)\Big];\\
&\chi=\frac{2e^2Z_0^2}{\Delta_{12}m_0};
\xi=\frac{3}{4}\frac{\epsilon_{F1}^0+\epsilon_{F2}^0}{\Delta_{12}}\chi+\frac{1}{2}\frac{2e^2(Z_2^2-Z_1^2)}{\Delta_{12}m_0}
\end{split} 
\label{df}
\end{equation}
, where $\epsilon_{Fi}^0$ is Fermi energy counted form the bottom of $i$-th subband at zero magnetic field. For the studied system $\epsilon_{F1}^0$=21.83 (meV); $\epsilon_{F2}^0$=6.68 (meV) and $\Delta_{12}$=15.15(meV) yield $\chi=1.12\cdot 10^{-5}[d(nm]^2$ and  $\xi=1.91\cdot 10^{-5}[d(nm]^2$. These  results are used to compute the parameter $X$ in Eq.(\ref{xpar}) up to terms proportional to $B_\parallel^4$.

\section{Appendix B}

The  expression (\ref{cond3}) contains energy integration of a product of two cosine functions.    To perform the integration we represent this product  as a sum of two cosines, oscillating at frequency $f_1+f_2$ and $\delta f=f_1-f_2$.  An integration of the cosine, oscillating at frequency $f_1+f_2$,  leads to an exponentially  small term $\sim \exp(-2\pi^2 (f_1+f_2) kT)$. Since $f_i kT \gg 1$  this term is neglected. 

To perform the integration in the $kT$ vicinity of Fermi energy $\epsilon_F$ we substitute $\epsilon=u+\epsilon_F$. After the substitution the phase of the second cosine, oscillating at frequency $\delta f$ is  a sum of two terms: $ \alpha=\pi\delta f u \sim u$ and $\beta=2\pi (f_2\Delta_g+\delta f \epsilon_F)$=const.  The cosine can be rewritten, using the identity: $\cos(\alpha+\beta)=\cos(\alpha)\cos(\beta)-\sin(\alpha)\sin(\beta)$. An integration of the product of two sine functions in the vicinity of the Fermi energy yields zero, since  $\sin(\alpha)$ is odd function of variable $u$, whereas $\partial f_T(u)/\partial u$ is even function of $u$. In result the integral is proportional to   $\langle \cos(2\pi \delta f u) \rangle \cos(2\pi f_2\Delta_g+2 \pi \delta f \epsilon_F)$. The integration vs $u$ yields:   $\langle \cos(2\pi \delta f u) \rangle= X/\sinh(X)$,   where $X=2\pi^2 kT \delta f$,\cite{ziman}  leading to Eq.(\ref{cond_fin}).

\end{document}